# Influence of substrate on structural and transport properties of LaNiO$_3$ thin films prepared by pulsed laser deposition

L. Cichetto Jr[1,2,3], S. Sergeenkov[4,*], J. C. C. A. Diaz[1], E. Longo[2,3] and F.M. Araújo-Moreira[1]

[1]Department of Physics, Universidade Federal de São Carlos, 13565-905 São Carlos, SP, Brazil
[2]LIEC - Department of Chemistry, Universidade Federal de São Carlos, 13565-905 São Carlos, SP, Brazil
[3]Institute of Chemistry, Universidade Estadual Paulista - UNESP, 14801-907 Araraquara, SP, Brazil
[4]Department of Physics, Universidade Federal da Paraíba, 58051-970 João Pessoa, PB, Brazil

**Abstract**

We report the structural and transport properties of LaNiO$_3$ thin films prepared by pulsed laser deposition technique. To understand the effects of film thickness, lattice mismatch and grain size on transport properties, various oriented substrates were used for deposition, including single-crystalline SrLaAlO$_4$ (001), SrTiO$_3$ (100) and LaAlO$_3$ (100). To achieve a high quality LaNiO$_3$ thin films, the vital parameters (such as laser fluence, substrate temperature, oxygen pressure, and deposition time) were optimized. The best quality films are found to be well textured samples with good crystalline properties.

**Keywords:** LaNiO$_3$, thin films, pulsed laser deposition, resistivity.

[*]Corresponding author; e-mail address: sergei@df.ufscar.br

1. **Introduction**

In recent years, the study of nonvolatile memories for use in storage and ferroelectric random access memory (FeRAM) devices has gained a new impetus. Manufacturing of these types of devices requires bottom electrodes in the form of thin films with some specific properties such as a high metallic conductivity and good adhesion to the substrates. Among various oxides, including $SrRuO_3$, $La_{1-x}Sr_xCoO_3$ and $LaNiO_3$ (LNO) [1-4] the use of perovskite-type conductive materials as bottom electrodes favors the growth of high quality ferroelectric thin films, such as BTO and PZT [5,6]. From the remarkable family of perovskite oxides, lanthanum nickelate LNO is a rare example characterized by metallic behavior in a wide range of temperatures being at the same time structurally compatible with many active functional layers [7-10]. LNO has attracted much attention due to its ability to drastically improve the ferroelectric fatigue problem [11,12]. Its pseudocubic lattice structure (with a cell parameter of 3.84 Å) is compatible with that of silicon (and even PZT) showing resistivity as small as 225µΩ cm at 300K [13]. Several studies have been reported on LNO thin films grown by different deposition techniques using $SrTiO_3$ and $LaAlO_3$ as popular substrates [14-16]. At the same time, very few systematical studies have been reported regarding deposition of LNO films on oriented $SrLaAlO_4$ substrates. The thermal expansion of single crystalline $SrLaAlO_4$ is around 7.38 ppm/°C at 462°C which is much less than that of other perovskite-type materials. Besides, the value of corresponding dielectric loss (~ $10^{-4}$) is equivalent to alumina [17] thus making this substrate especially attractive for manufacturing FeRAM devices. A rather significant advancement of technology in recent years made it possible to fabricate devices on the nanometer scale. The main idea of our study is to investigate to what extent we can miniaturize (with respect to film thickness) the conductive layers of LNO maintaining good properties of its metallic behavior.

In this paper we present our results on fabrication and systematic study on structural characterization and transport properties of LNO thin films grown by pulsed laser deposition (PLD) technique on highly oriented substrates, $SrLaAlO_4$, $SrTiO_3$ and $LaAlO_3$. The structure of thin films deposited on these substrates was characterized by X-ray diffraction (XRD), field emission scanning electron microscopy (FE-SEM), and atomic force microscopy (AFM). The electrical resistivity measurements as a function of temperature were used to verify and confirm the metallic behavior of our films.

2. **Experimental**

In order to provide high quality samples, PLD technique was used for deposition of thin films. When using the PLD technique for manufacturing thin films one should first obtain a target with high enough density because a compact and uniformly dense target is required to produce a laser plume with good quality and to have a consistent ablation of the surface of the target. The dense and crack-free LNO circular target with diameter of 5*cm* and thickness of 1.25*cm* was prepared from highly pure polymeric precursors by Pechini method [18] using $La_2(CO_3)_3 \times H_2O$ (99.9% Aldrich) and $Ni(OCOCH_3)_2 \times 4H_2O$ (98% Aldrich). The calcination and sintering were performed in the air at 900°C for 4*h* and at 1200°C for 6*h*, respectively. The thin films were grown on $SrLaAlO_4$ (SLAO), $SrTiO_3$ (STO) and $LaAlO_3$ (LAO) substrates by PLD technique using a KrF excimer laser with 248 nm wavelength and 25 ns pulse duration and deposition rate of 4 Hz. The laser beam was focused at the angle of 45° on target of LNO with fluence ranging from 1.5 to 1.8 J/cm² depending on the type of substrate used. To reduce non-uniform erosion, the target was rotated during the ablation process and at the end of each deposition the target was sanded and polished before the next deposition. The distance between the target and the substrate ranged from 4.5 to 5 cm with deposition temperature between 610°C and 630°C. The temperature was controlled by a thermocouple placed immediately at the back side of the substrate. Before the start of the deposition, a base pressure of $P_{base} = 5 \times 10^{-8}$ *mbar* was achieved in the deposition chamber using a turbo molecular pump. Thereafter, oxygen was injected into the deposition chamber monitored by the mass flow control with values between 80 and 95 SCCM depending on the substrate used for deposition of thin films. In order to achieve preferred crystallographic orientation and good electrical conductivity, the oxygen deposition pressure of $P_{dep} = 0.2$ *mbar* was used. Post deposition in-situ annealing carried oxygen at a pressure of $P_{oxy} = 5 \times 10^2$ *mbar* for *1h* to avoid formation of oxygen vacancies and maintain the phase stoichiometry of the films. This procedure is necessary to further improve the electrical conductivity of the LNO films. The values of the deposition parameters are summarized in Table 1, where $d_{t-s}$ is a distance between the substrate and the target, $T_{dep}$ is a deposition temperature, $\Phi$ is the fluence given for a laser spot with dimensions of 1mm x 6mm and $P_{dep}$ is a deposition pressure. To study the effects due to film thickness, samples were manufactured with thickness ranging from 25 nm to 55 nm which has been controlled by the number of laser shots.

## 3. Results and Discussion

XRD measurements of the films were carried out using a Shimadzu XRD-6000 diffractometer with CuKα radiation. The unit cell refinements were performed using the Le Bail method through the GSAS/EXPGUI code [19,20]. The results shown in Fig.1(a) revealed that, independent of the substrate, our films crystallize into a cubic perovskite structure with a space group symmetry Pm3m (221). For different substrates a similar behavior in the lattice parameter with the variation of the thickness was observed. In general, we may conclude that the evolution of the lattice parameters is inversely proportional to the thickness of the films with a clear tendency toward the bulk lattice parameter for greater thicknesses. According to Fig.1(b), the relative variation of the lattice parameters of our films is no higher than 0.1Å for each substrate with the least variation (around 0.03Å) observed in the films grown on SLAO. Also, it was confirmed that the growth of the films occurs in the (100) direction. The thicknesses of all films were measured by FEG-SEM (cross-sectional area) and their values were used for calculation of resistivity. Typical FEG-SEM images for LNO/STO, LNO/LAO and LNO/SLAO structures are shown in Fig.2(*a*)–(*c*). An image of silver contacts with a diameter of 0.5 mm used for electric measurements is depicted in Fig.2(*d*). A silver pad was evaporated onto the surface of the heterostructure to form the top electrode. Several films have been prepared for this study but only the films with the thicknesses of 25, 35, 45 and 55nm were taken into account. The analysis based on FE-SEM images revealed no separation between LNO films and substrates, thus indicating a good adherence between the film and the substrate. Fig.2 depicts some typical FE-SEM images of LNO thin films deposited on (*a*) STO substrate, (*b*) LAO substrate, and (*c*) SLAO substrate; (*d*) circular silver contacts with diameter of 0.5 mm used for electric measurements. The resistivity measurements as a function of temperature were performed using the four-point method on as-deposited LNO thin film. The probed temperature region is ranging from 10 to 300 K. Fig.3(*a-c*) presents typical resistivity data. The temperature dependence of the derivative *dρ/dT* shows a good metallic behavior for all samples. Fig.3(d) exhibits values of resistivity taken at 300 K compared to the thicknesses of thin films deposited on different substrates. The observed behavior for different types of substrates (an increase of the values of resistivity with decreasing the thickness of thin films) is quite similar to previously reported results [8-10]. However, the LNO/STO heterostructure with thickness of 35 nm exhibits unusually small values of resistivity in comparison with previous observations [15,21]. The surface morphology of the films was checked by using atomic force microscopy (AFM). More precisely, the film roughness is defined by the root mean square (RMS) value. Typical AFM images for LNO films deposited on different substrates are shown in Fig.4(*a*)–(*c*). Measurements were made over the projected area of 1μm x 1μm. A much lower roughness (below 1.5 nm) has been observed in comparison with the results of the other groups [22-25] making our films more suitable for their practical applications. Fig.4(*d-f*) represents the dependence of the average grain size (left axis) and RMS roughness (right axis) on the thickness of the films deposited on different substrates. The observed increase of RMS roughness with the increase of film thickness is probably due to a larger grain size formation as well as to an increase in the porosity of the films.

4. **Conclusion**

In summary, highly oriented LNO thin films have been successfully fabricated using PLD technique on various substrates ($SrLaAlO_4$, $SrTiO_3$ and $LaAlO_3$) with different thickness (25, 35, 45 and 55 nm). The structure probing methods (including FEG-SEM, XRD and AFM) and electrical measurements were used to determine the effect of crystallinity, orientation, and surface morphology of LNO thin films. The results obtained by electrical measurements indicate that LNO is a good conductor. AFM data show that the film roughness becomes larger with increasing the thickness. Based on our findings, we may conclude that the epitaxial LNO thin films deposited on oriented SLAO (001) substrates are very attractive for their use as bottom electrodes in the production of memory type devices.

**Acknowledgments**

We are indebted to Marcel Ausloos (Liege) and Alex Kuklin (Dubna) for useful discussions. We are very grateful to NanO LaB for their help with resistivity measurements. We would like to thank LMA-IQ for allowing us to use FEG-SEM facilities. This work was financially supported by Brazilian agencies FAPESP, FAPESQ (DCR-PB) and CNPq. We are very thankful to FAPESP (CEPID CDMF 2013/07296-2 and 2014/01371-5) for continuous support of our project on nickelates.

**Figure Captions**

**Fig.1.** (Color online) (a) XRD patterns for LNO films deposited on different substrates; (b) relative variation of the lattice parameters with the thickness of the films.

**Fig.2.** (Color online) Typical FE-SEM images of LNO thin films deposited on (*a*) STO substrate, (*b*) LAO substrate, and (*c*) SLAO substrate; (*d*) circular silver contacts with diameter of 0.5 mm used for electric measurements.

**Fig.3.** (Color online) The dependence of the measured resistivity $\rho$ on temperature for LNO films deposited on (*a*) STO substrates, (*b*) LAO substrates, and (*c*) SLAO substrates; (*d*) values of resistivity taken at 300 K compared to the thicknesses of thin films deposited on different substrates.

**Fig.4.** (Color online) Typical AFM images for LNO films deposited on (*a*) STO substrates, (*b*) LAO substrates, and (*c*) SLAO substrates; (d-f) dependence of the average grain size (left axis) and RMS roughness (right axis) on the thickness of the films deposited on different substrates (a-c).

**Table 1**: Optimized PLD parameters for bottom electrode of LaNiO$_3$ thin films for different types of substrates.

| Substrate | $d_{t-s}$ (cm) | $T_{dep}$ (°C) | $\Phi$ (J/cm$^2$) | $P_{dep}$ (mbar) |
|---|---|---|---|---|
| **SrLaAlO$_4$** | 4.5 | 610 | 1.50 | 0.20 |
| **SrTiO$_3$** | 5.0 | 625 | 1.80 | 0.21 |
| **LaAlO$_3$** | 5.0 | 630 | 1.75 | 0.21 |

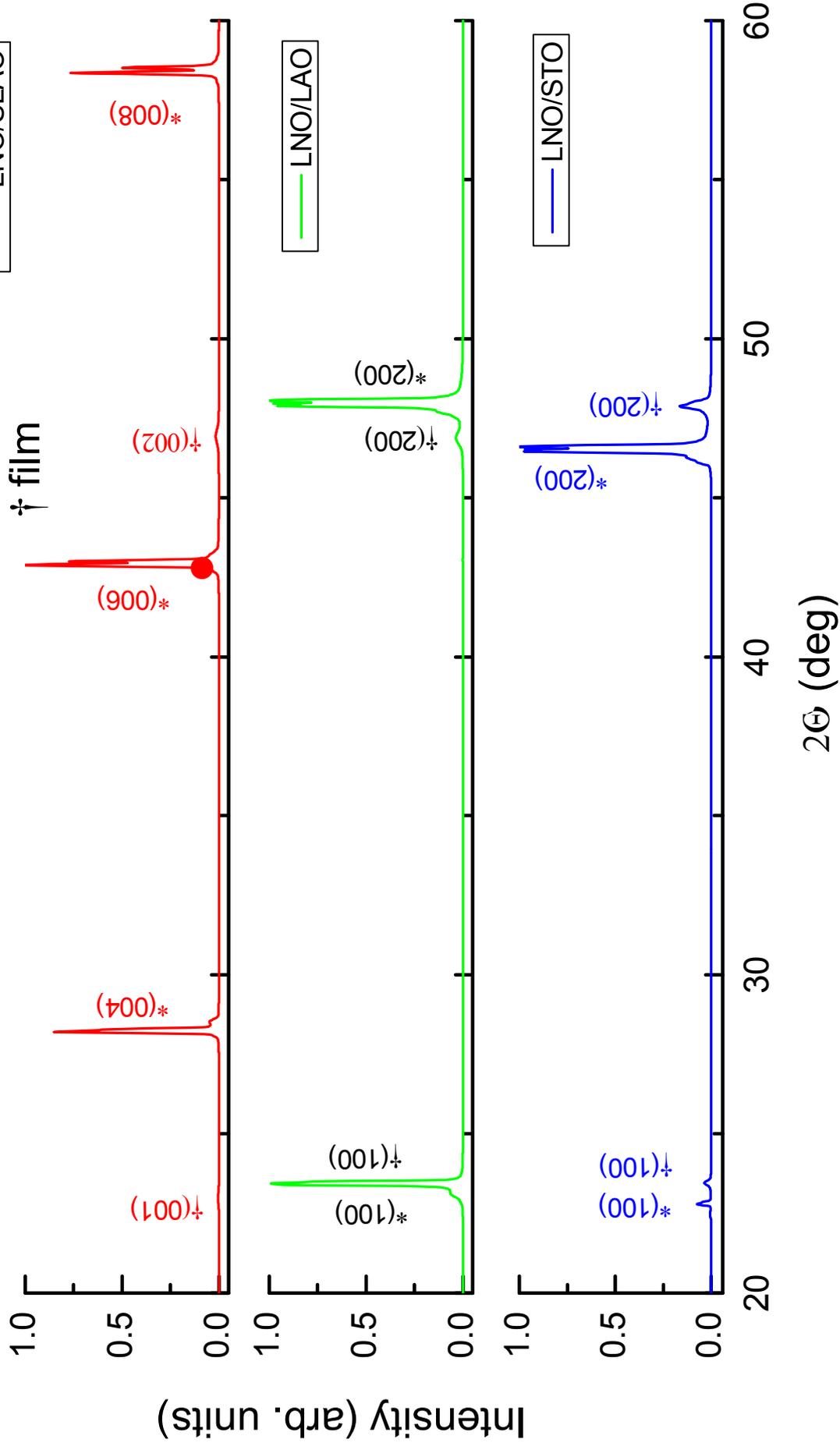

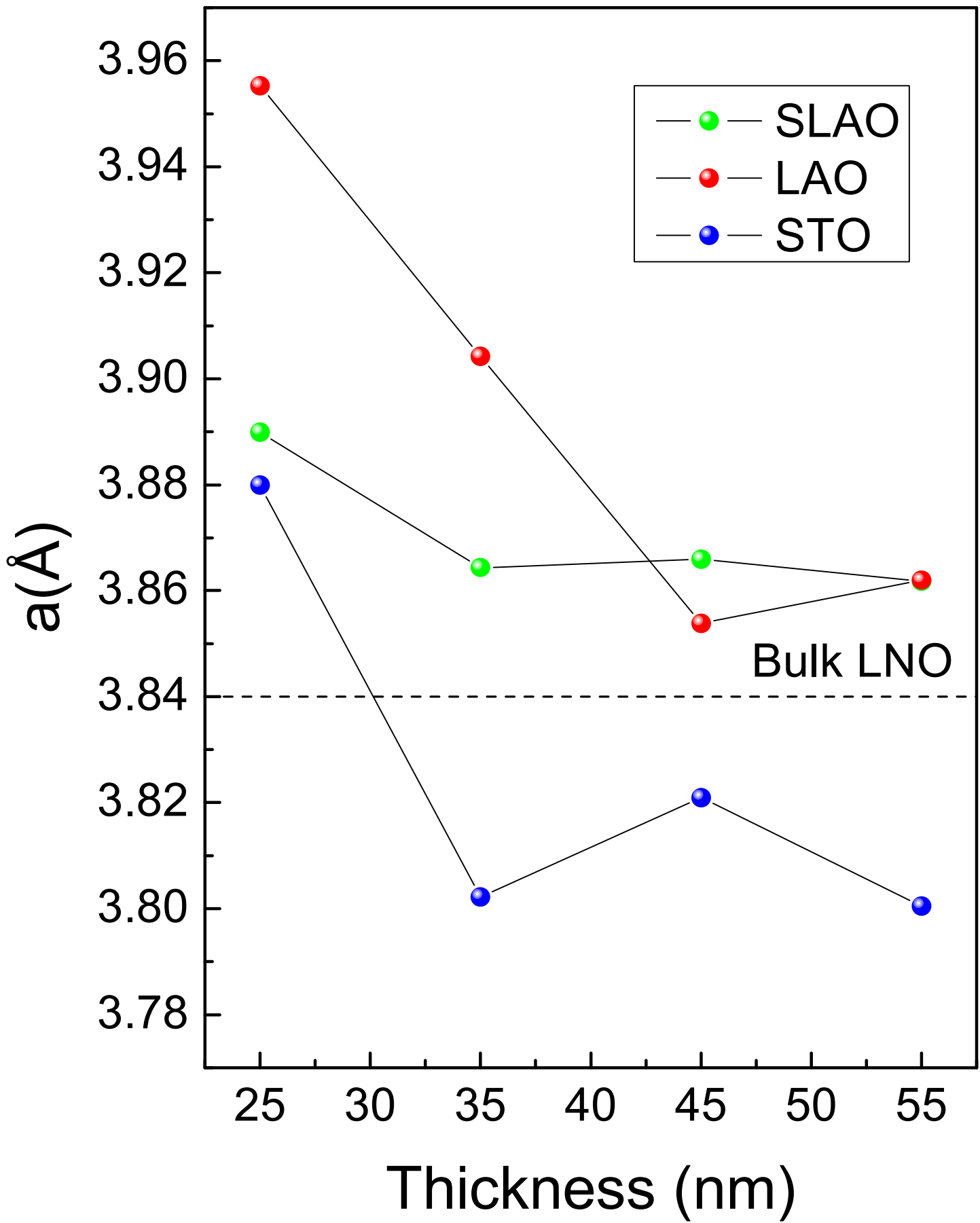

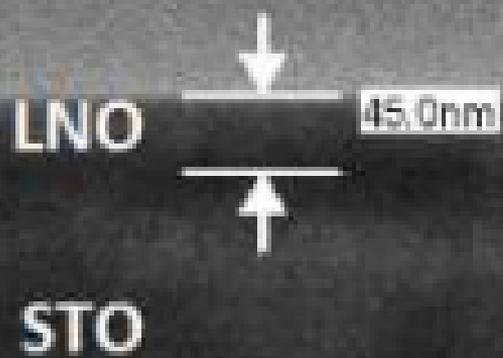

LNO

LAO

55.0nm

X 150,000  3.0kV  SEI  SEM  WD 6.8mm  100nm

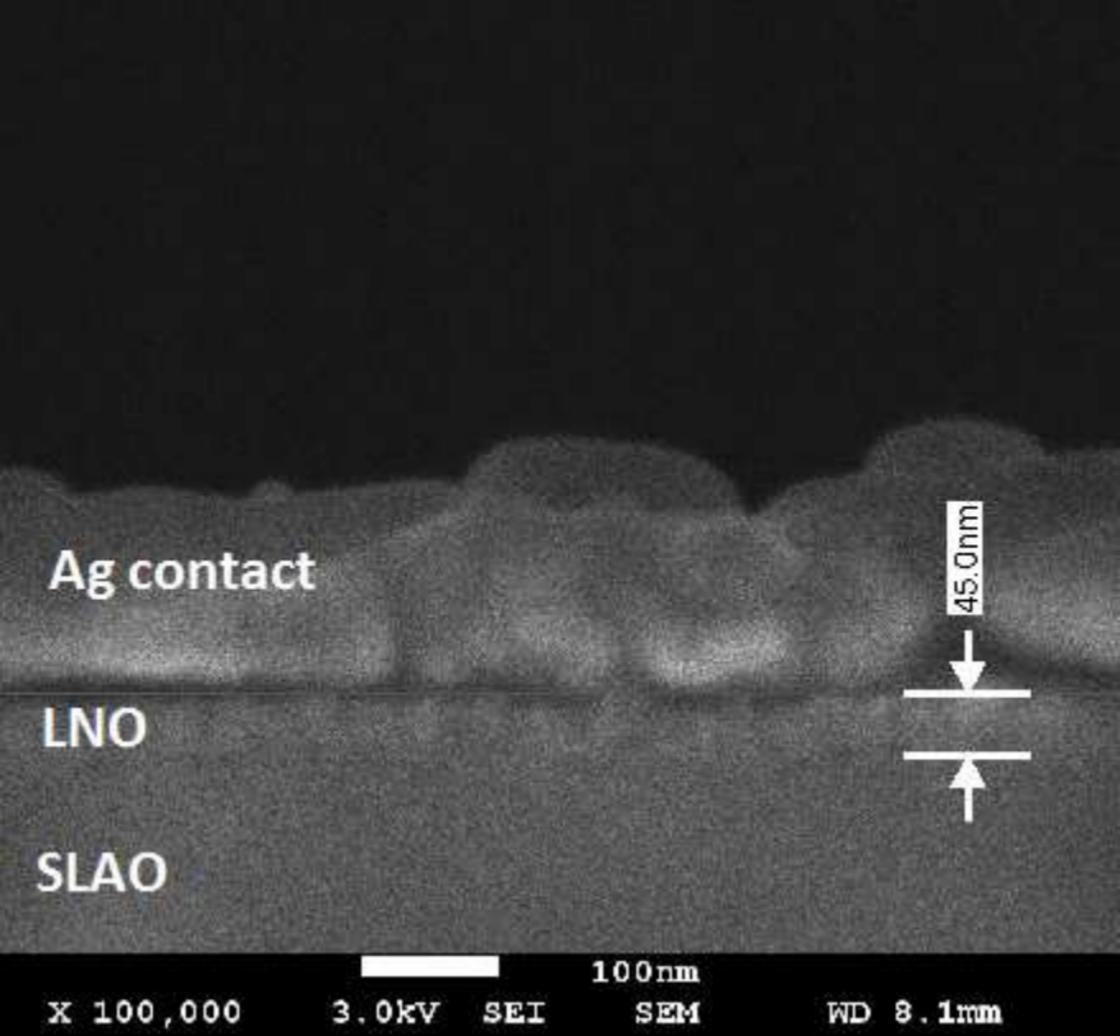

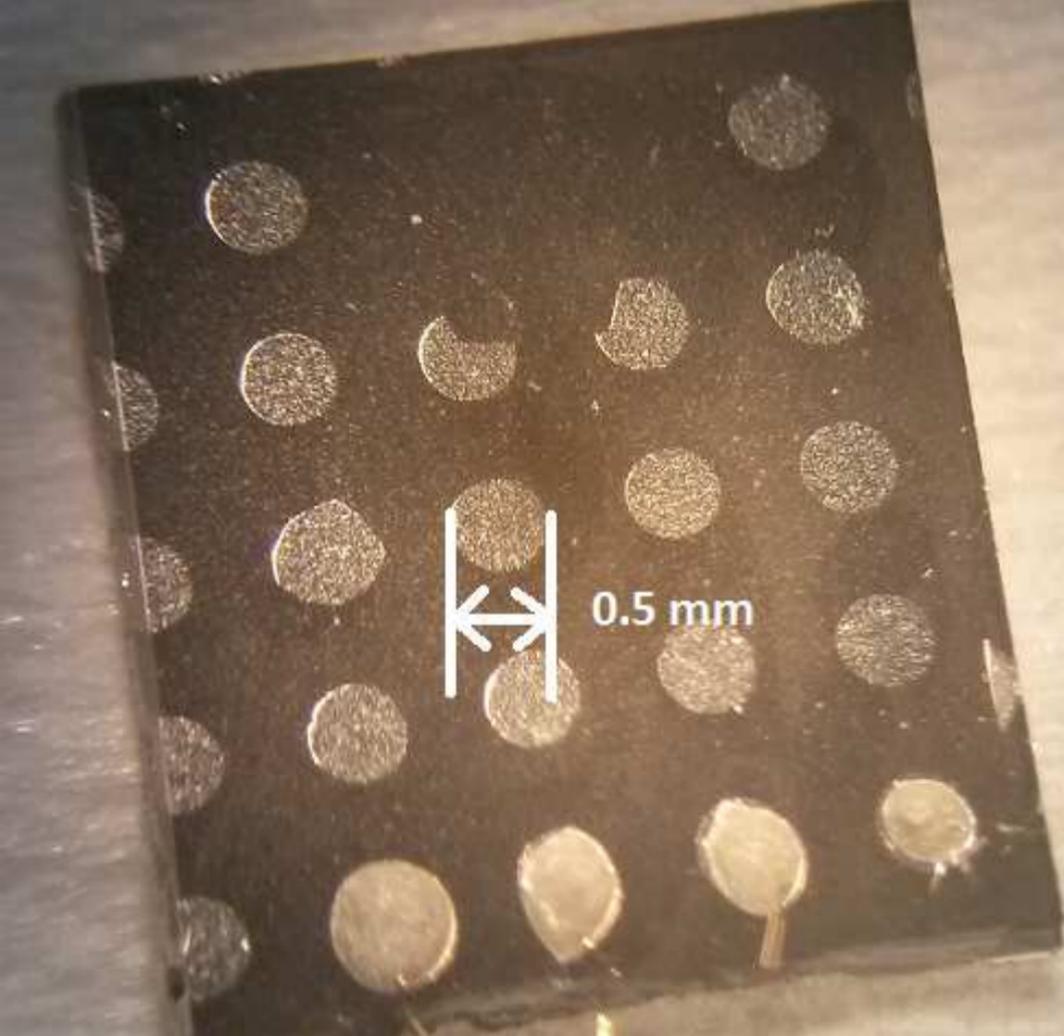

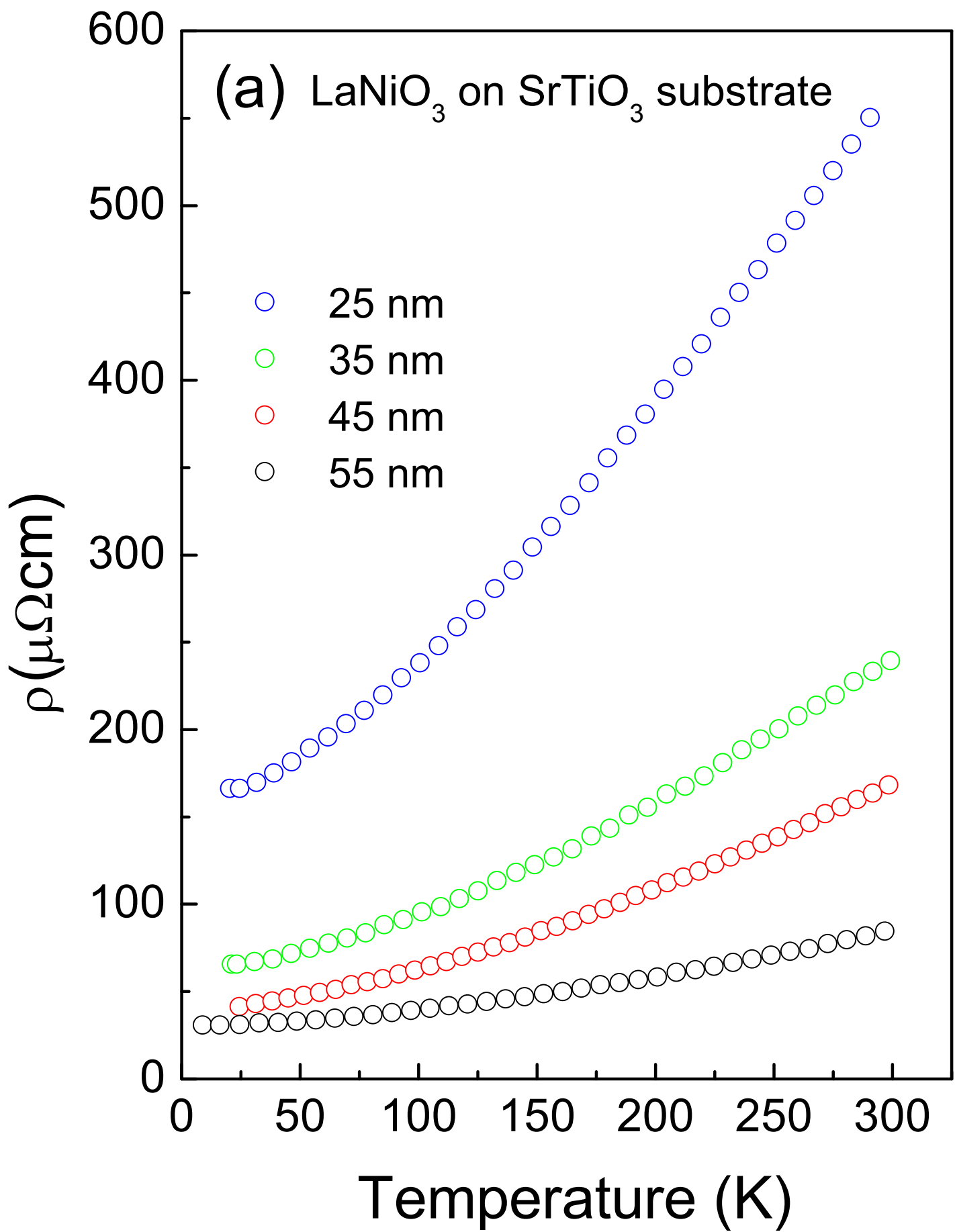

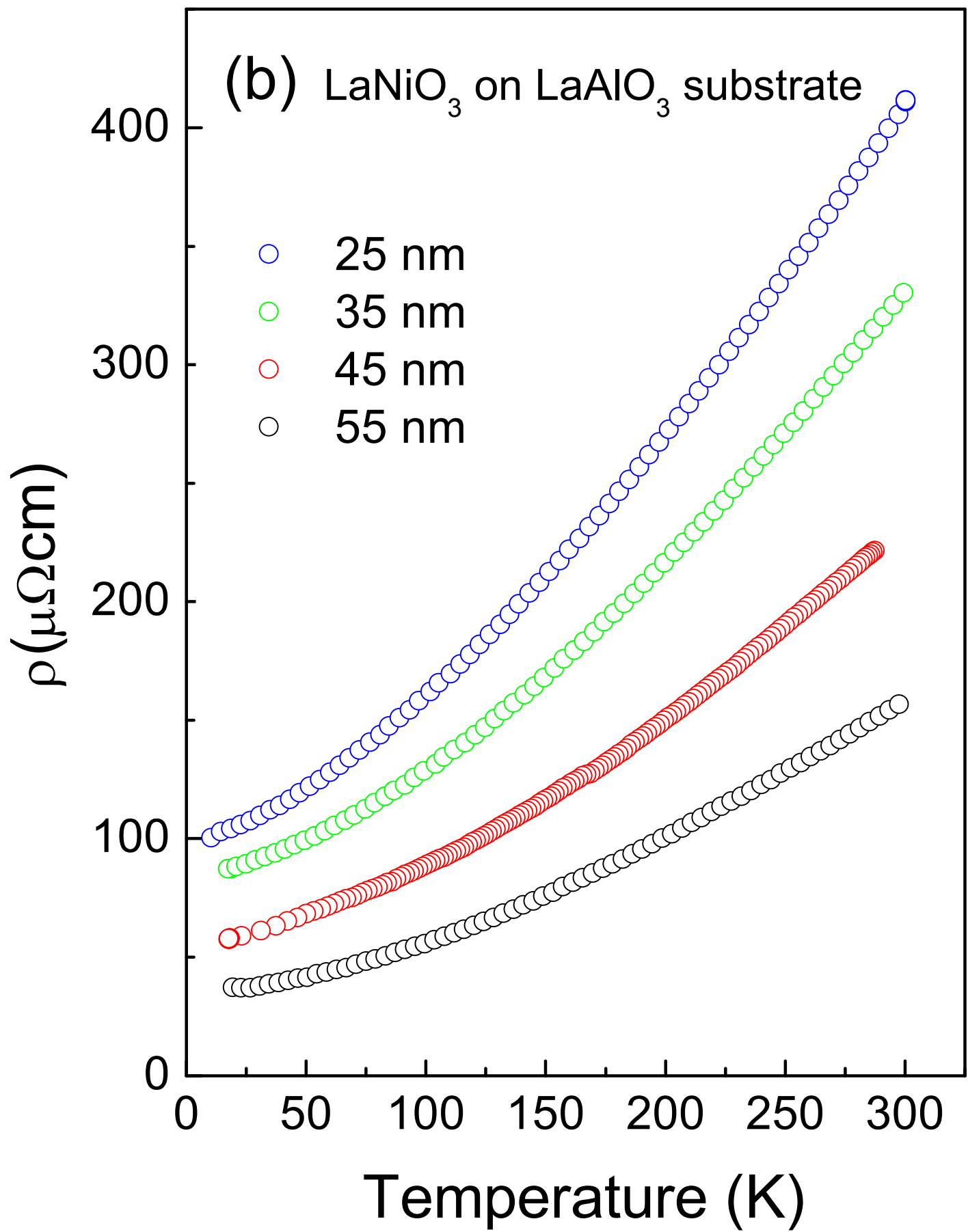

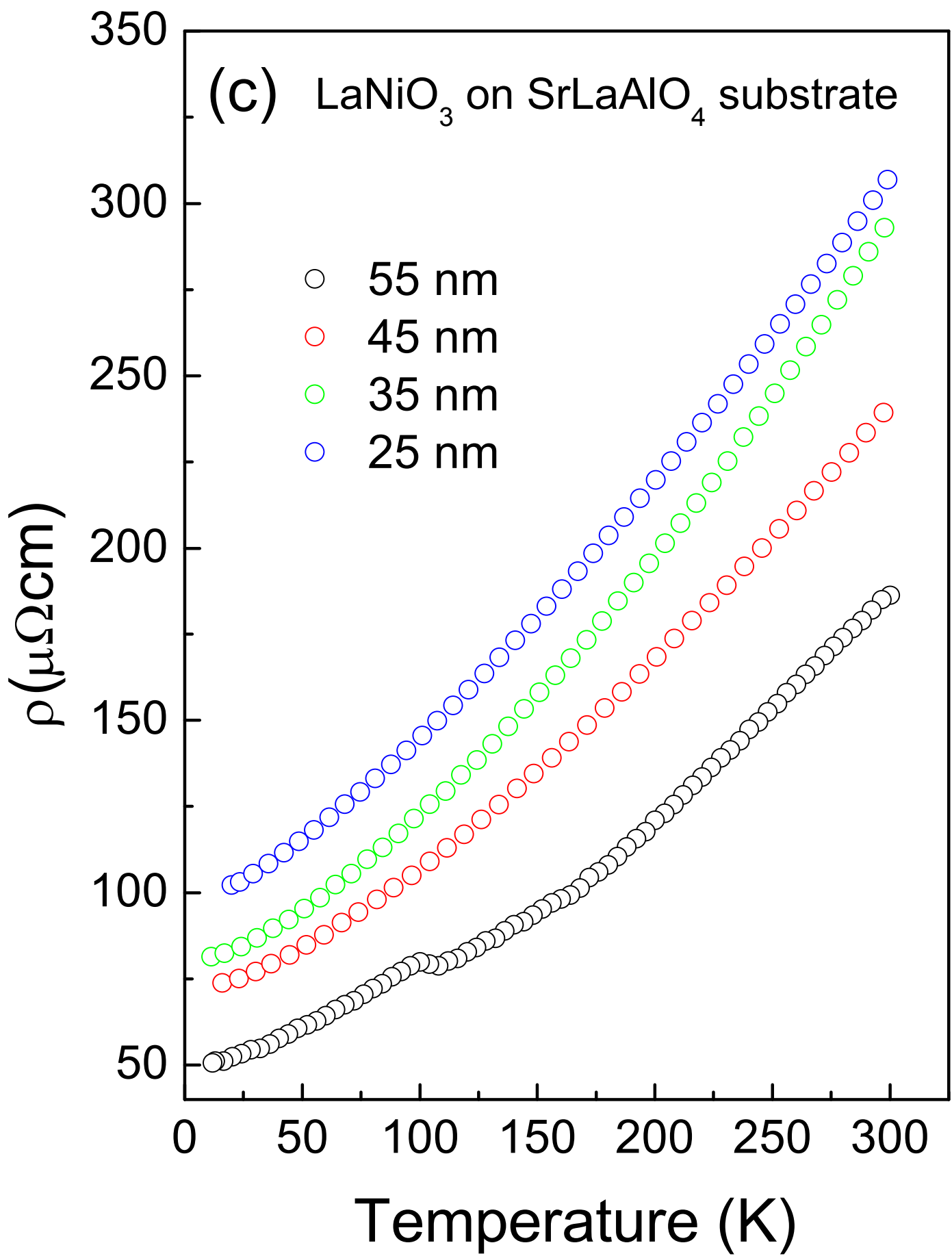

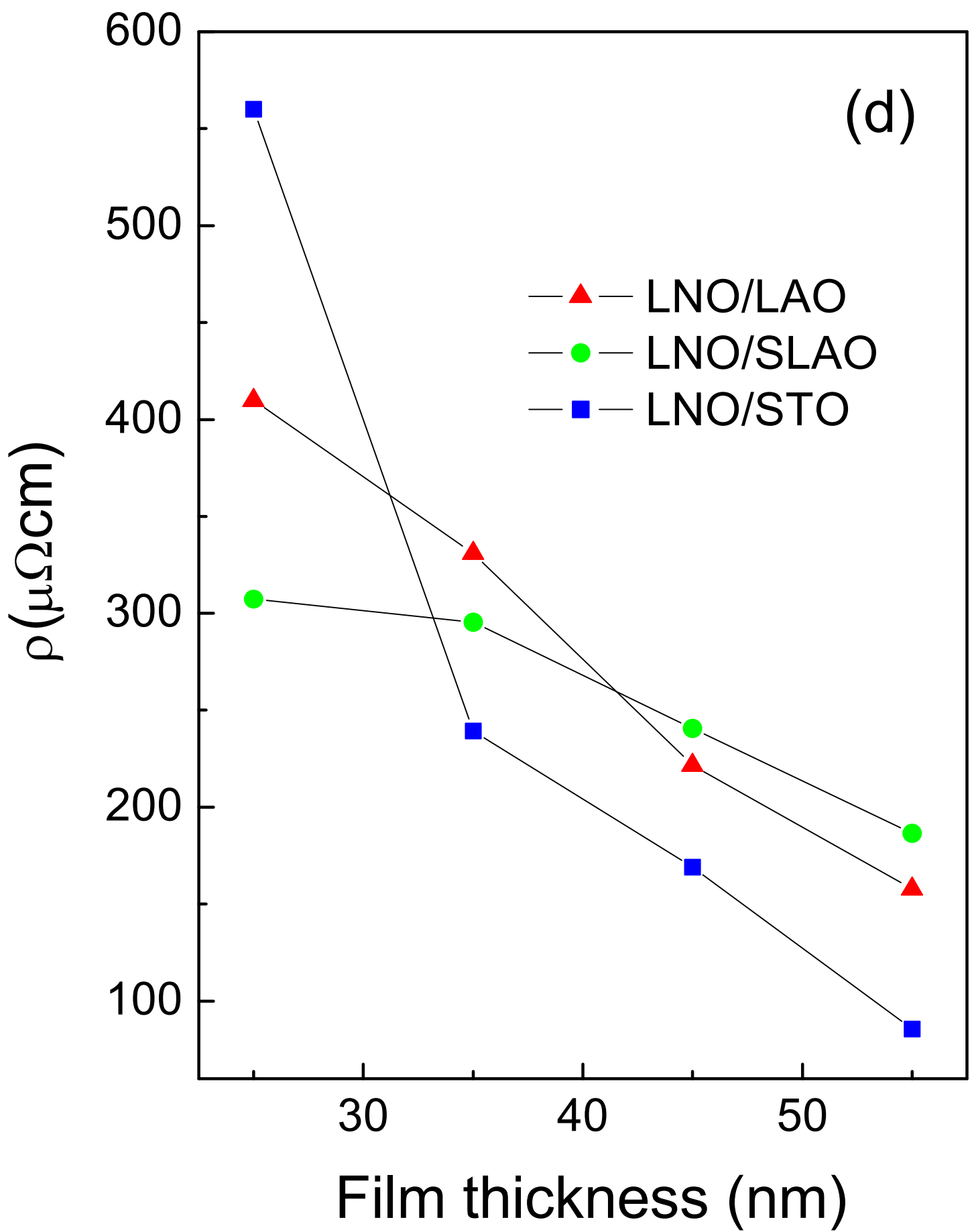

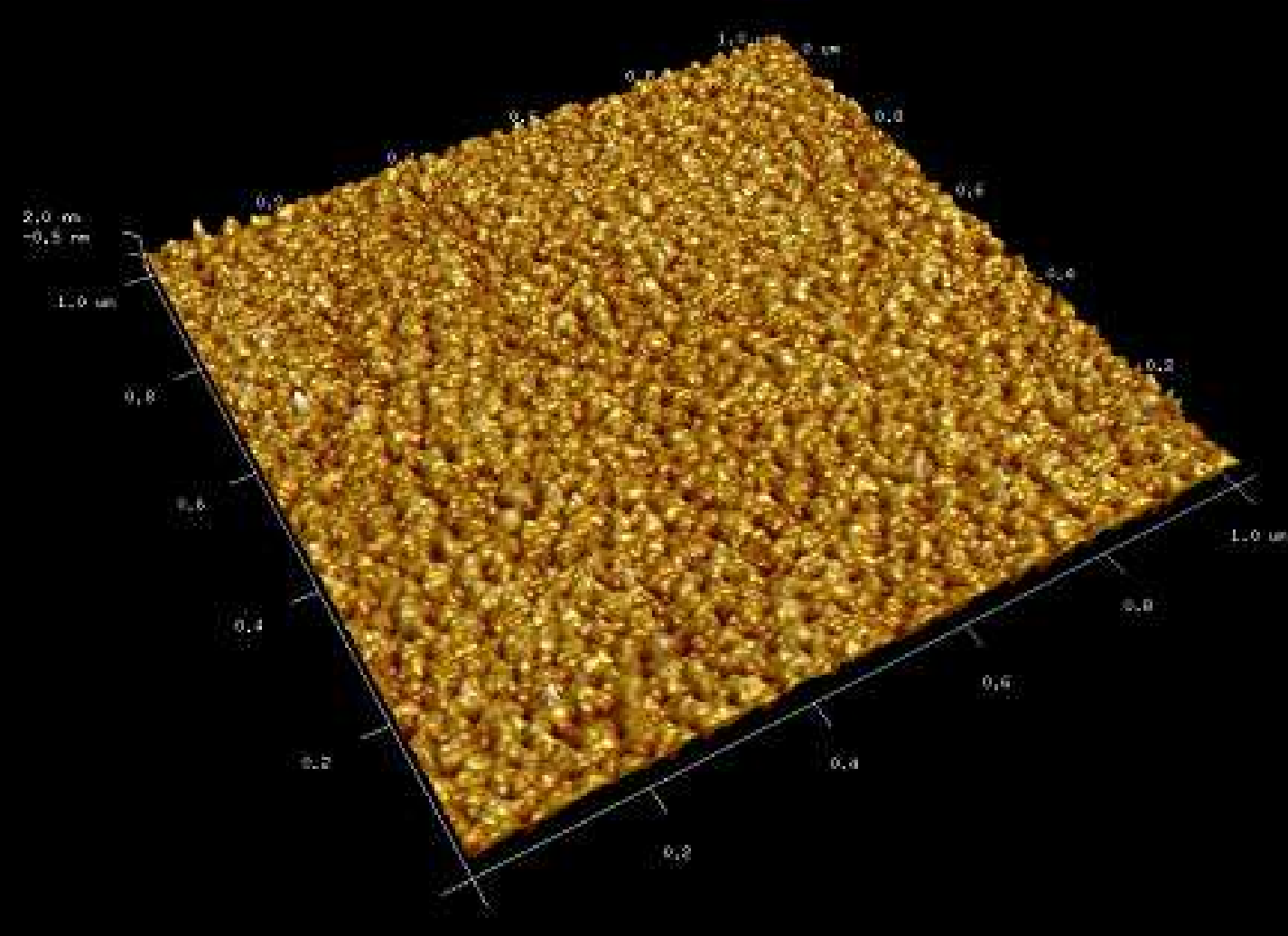

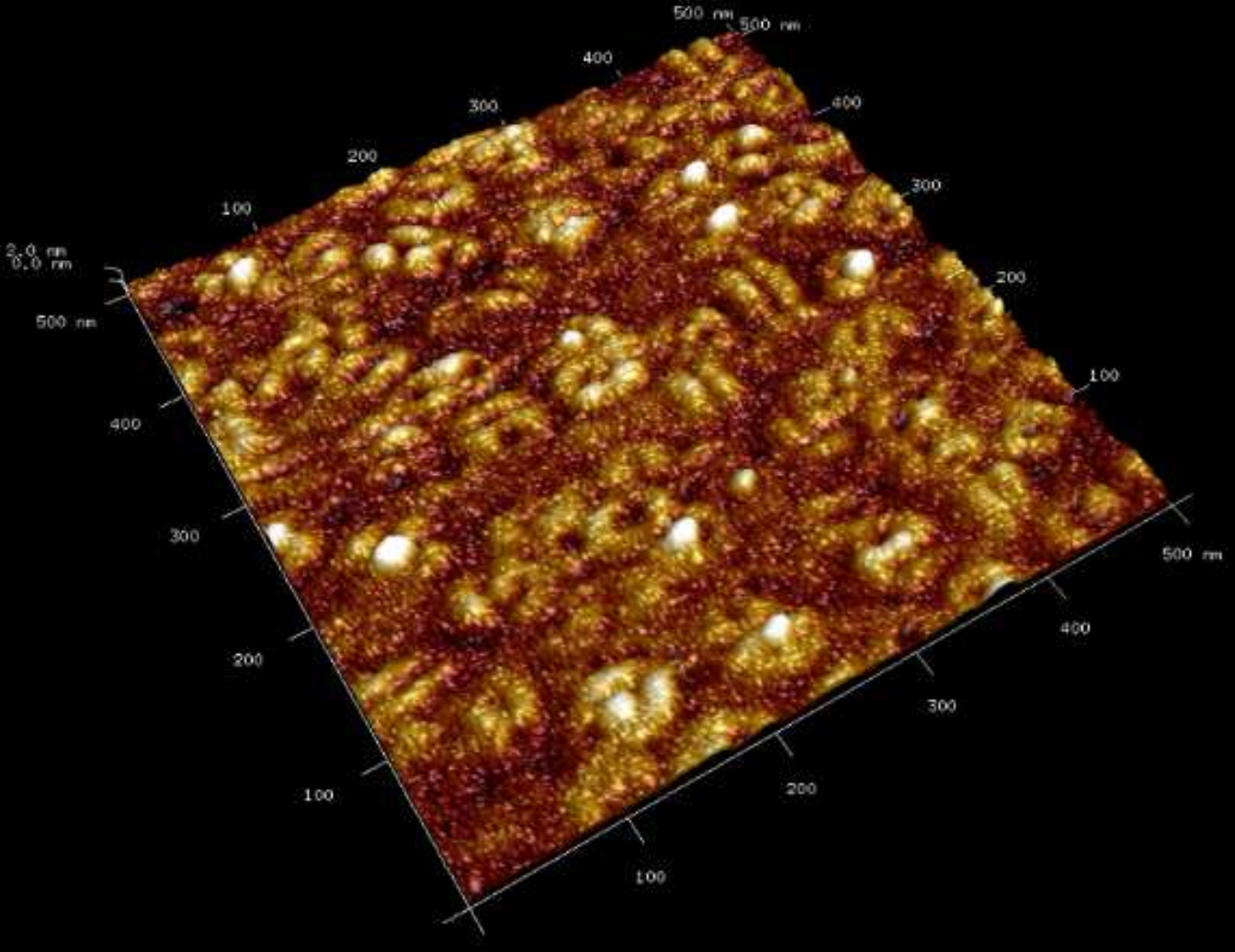

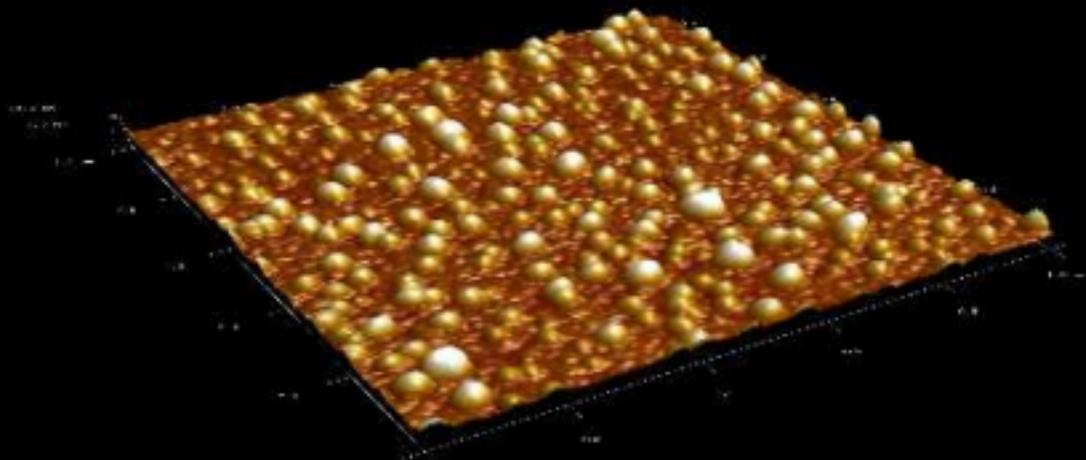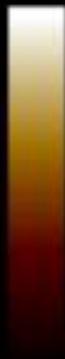

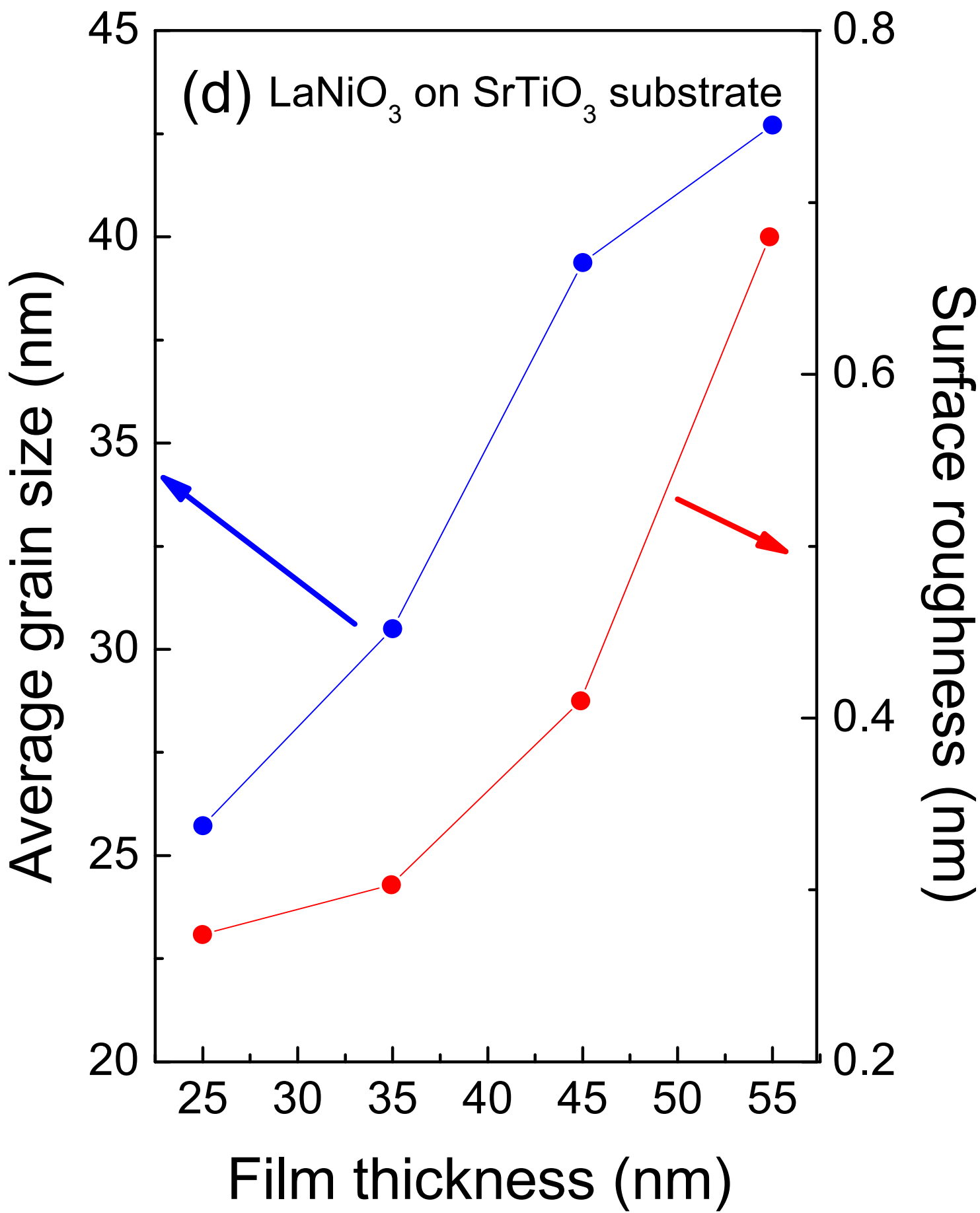

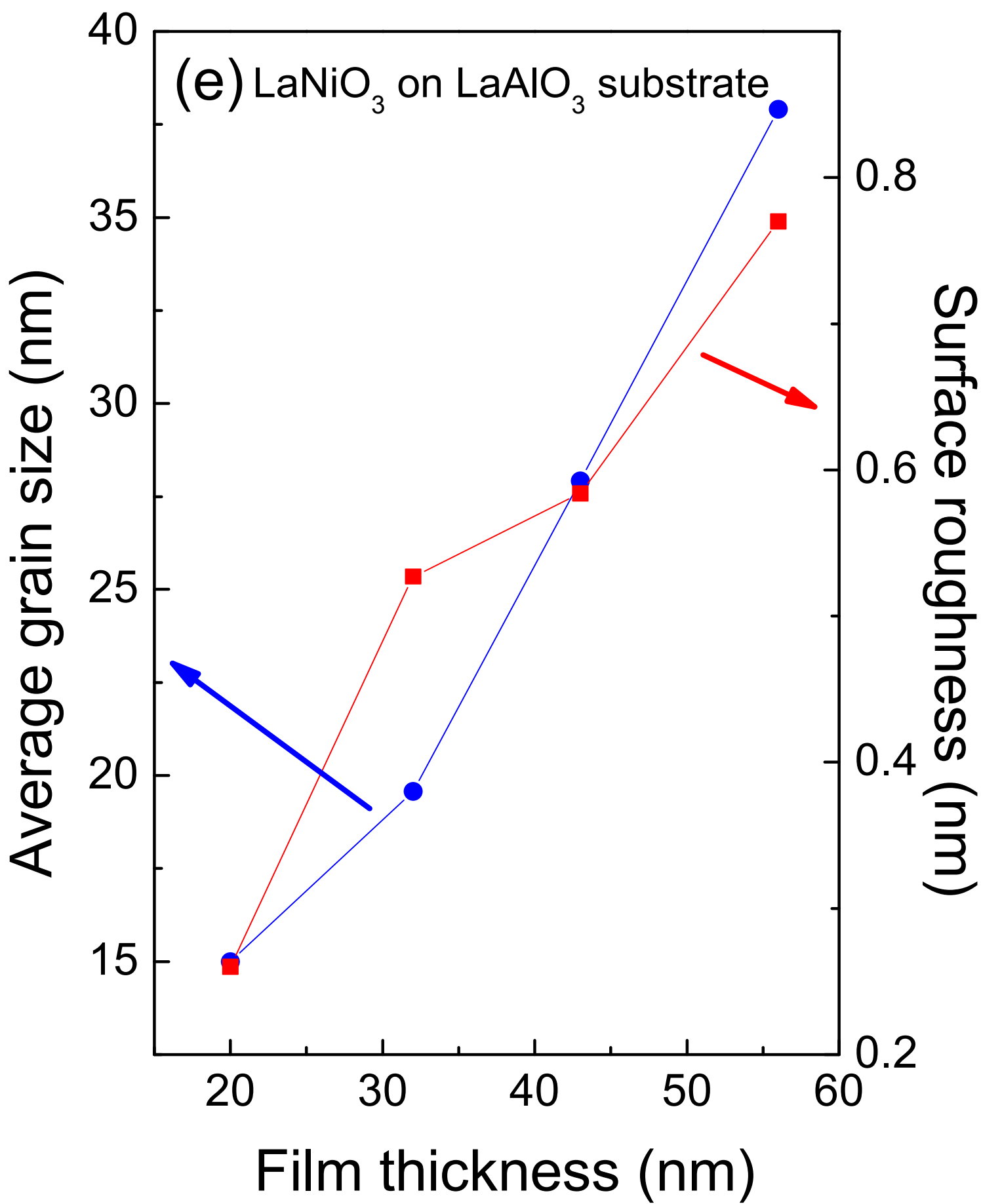

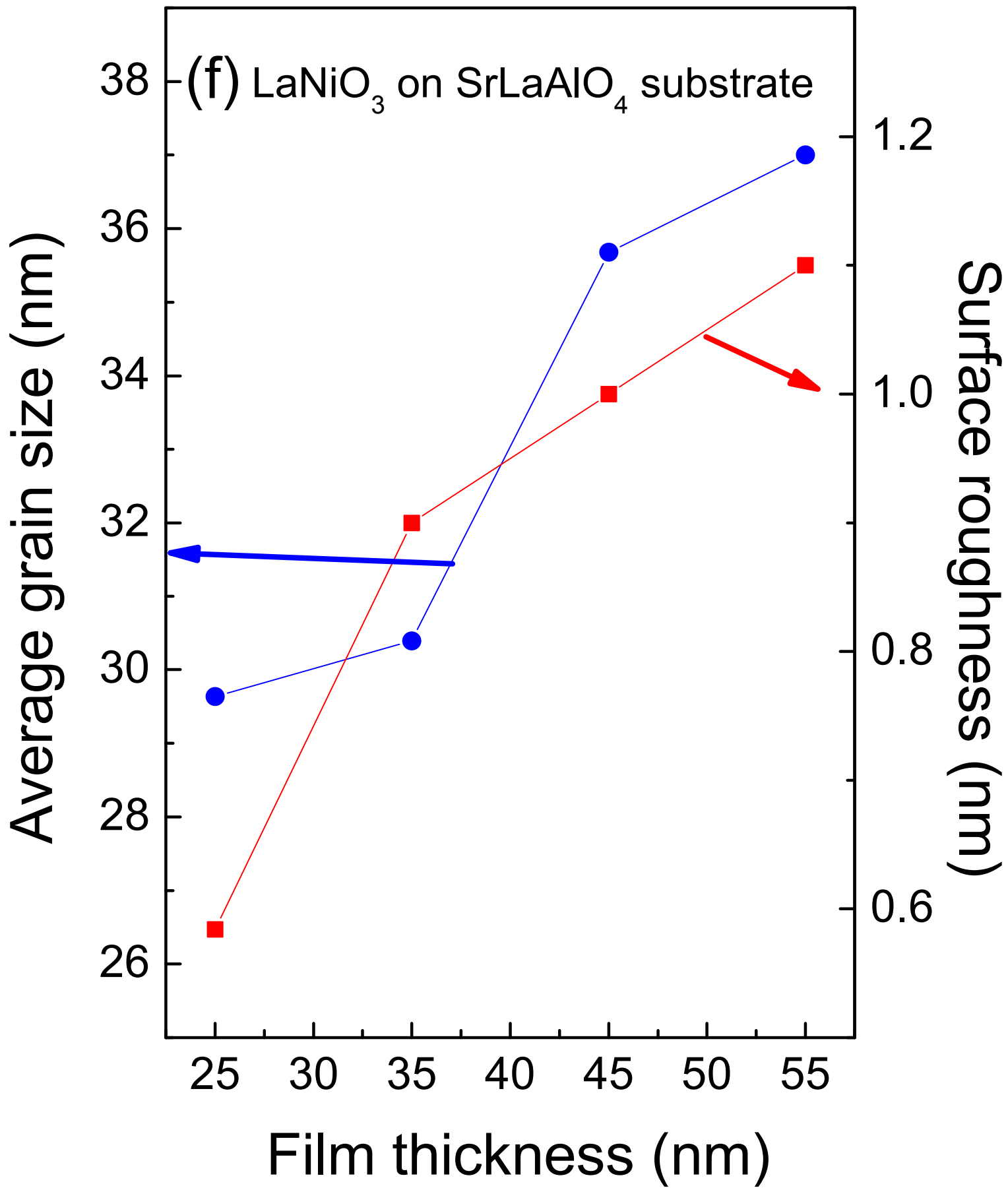